\documentclass[]{aastex}
\usepackage{emulateapj5}

\usepackage[dvips]{epsfig}

\begin{document}

\title{The Jet and the Supernova in GRB~990712\altaffilmark{1}}
\altaffiltext{1}{Based on observations made with ESO Telescopes at La Silla 
(ESO Programme 59.A-000) and Paranal, Chile (ESO Programme 63.O-0567(B)),
and on observations with the NASA/ESA {\em Hubble Space Telescope}, obtained from 
the data archive at the Space Telescope Institute, which is operated by the 
Association of Universities for Research in Astronomy Inc.\ under contract
NAS5-26555}

\author{G.~Bj\"ornsson\altaffilmark{2}, J.~Hjorth\altaffilmark{3}, 
P.~Jakobsson\altaffilmark{2,3}, L.~Christensen\altaffilmark{3}, 
S.~Holland\altaffilmark{4,5}}

\altaffiltext{2}{Science Institute, University of Iceland, Dunhaga~3, 
  IS--107 Reykjavik, Iceland, e-mail:gulli@raunvis.hi.is}
\altaffiltext{3}{Astronomical Observatory, University of Copenhagen, 
  Juliane Maries Vej 30, DK--2100 Copenhagen \O, Denmark}
\altaffiltext{4}{Department of Physics, University of Notre Dame, Notre Dame,
  IN 46556-5670, U.S.A.}
\altaffiltext{5}{Institut for Fysik og Astronomi (IFA), Aarhus Universitet, 
 Ny Munkegade, Bygning 520, DK--8000 {\AA}rhus C, Denmark}

\begin{abstract}
The optical light curve of the afterglow following the gamma-ray burst 
GRB~990712 is re-examined. Recently published polarization measurements of 
that source require a collimated outflow geometry that in turn predicts
a break in the light curve. We show that the V-band light curve is consistent
with such a break and that the post-break light curve evolution is dominated 
by a supernova contribution.
\end{abstract}



\section{Introduction}
\label{sec:intro}

Optical light curves of gamma-ray burst (GRB) after\-glows decay as a power law 
in time, $F\propto t^\alpha$, with a typical value of the decay index $\alpha\sim -1$. 
In several cases the light curve has been observed to steepen, about 1--3 days after 
the gamma-ray event, to $\alpha\sim -2$ or even steeper (e.g.\ Kulkarni et~al.\ 1999; 
Castro-Tirado et~al.\ 1999; \ Harrison et~al.\ 1999; Stanek et~al.\ 1999; 
Israel et~al.\ 1999; Holland et~al.\ 2001a; Jaunsen et~al.\ 2000; Jensen et~al.\ 2000). 
Such a light curve is commonly referred to as a broken power law with $\alpha_1$ 
denoting the pre-break decay index and $\alpha_2$ the post-break index.

A generic model that has been successfully applied to after\-glow observations is 
that of a relativistic fireball that sweeps up ambient matter and decelerates. 
An unbroken light curve can be explained by a spherically symmetric fireball 
(e.g.\ Sari, Piran \& Narayan 1998), whereas a broken power law in most cases 
requires a collimated outflow, i.e.\ a jet (Rhoads 1999; Sari, Piran \& Halpern 1999). 
In the latter case the light curve steepens as the relativistic beaming angle 
($\sim 1/\Gamma$, with $\Gamma$ the decreasing bulk Lorentz factor), increases 
and becomes equal to or greater than the jet opening angle, $\theta$.

The currently favored model for long-duration GRB progenitors is that 
of a collapsar (Woosley 1993), or hypernova (Paczy\'nski 1998). Numerical 
simulations show that as an iron core of a massive star collapses to form 
a black hole it releases up to $10^{52}-10^{53}$ ergs of energy, a fraction 
of which produces a jet and a gamma-ray burst. The remaining energy explodes 
the star and produces a supernova (MacFadyen \& Woosley 1999).  

The presence of a supernova is most easily confirmed by studying afterglow light 
curves. After an initial power law decay of the emission originating in the jet, 
an underlying supernova is expected to dominate the late time light curve behavior, 
in most cases appearing a few days to a couple of weeks after the gamma-ray event. 

A number of afterglows have been interpreted by such a scenario, 
e.g.\ GRB~980425/SN1998bw, an unusual Type Ib/c supernova located relatively 
nearby, at a redshift of $z=0.0085$ (Galama et~al.\ 1998). Other cases include 
GRB~980326 (Bloom et~al.\ 1999), GRB~970228 (Reichart 1999; Galama et~al.\ 2000), 
GRB~000418 (Dar \& De R\'ujula 2000), and possibly GRB~970514 (Germany et al.\ 2000; 
Turatto et~al.\ 2000 and GRB~980703 (Holland et~al.\ 2001b).

One counterexample may be GRB~990712, that apparently did not show a steepening 
light curve nor a supernova-like component. The optical discovery and early light 
curve of GRB~990712 was reported by Sahu et~al.\ (2000; hereafter referred to 
as S00), who found that a decay proportional to $t^{-1}$, plus a constant host 
contribution provides a better fit to the data than a $t^{-1}$ power law with 
a constant host and a supernova of type SN1998bw at the appropriate redshift.
Hjorth et~al.\ (2000) discussed the late afterglow properties as well as
the host galaxy. Their localization of the gamma-ray burst within the 
host was based on astrometric data adopted from S00 and turned out to be
incorrect. They concluded that no SN component was needed.
Fruchter et~al.\ (2000) correctly identified the burst location within the 
host from its variability.

A set of polarization measurements for GRB~990712, presented by
Rol et~al.\ (2000), showed a variable degree of polarization at a constant
position angle over a 24 h interval starting about 11 h after the 
gamma-ray event. Because of the constant position angle, Rol et~al.\ (2000)
concluded that none of the currently available models could explain the
observations. Bj\"ornsson \& Lindfors (2000; hereafter BL00) on the other hand, 
showed that 
the polarization data is most naturally explained by a collimated outflow
that was modestly spreading during the polarization measurements. They 
estimated the jet opening angle, $\theta$, to be about $6^\circ$. A consequence 
of that interpretation is that a break should appear in the light curve about
1--2 days after the gamma-ray event, for the same reason as the break in the 
light curves of GRB~990123 (e.g. Kulkarni et~al.\ 1999; Castro-Tirado et~al.\ 1999) 
and GRB~990510 (e.g.\ Harrison et~al.\ 1999; Stanek et~al.\ 1999; 
Israel et~al.\ 1999; Holland et~al.\ 2001a).

Prompted by the BL00 prediction, we have reanalyzed the $V$ and $R$ band light curves 
of GRB~990712. We show that a break indeed appears to be present in the $V$-band at 
the time predicted by BL00. As a consequence, a prominent supernova-like component 
appears in the post-break light curve that is also clearly observed in the $R$-band
where no sign of a break is detected. The data provides a tantalizing 
case for the GRB/SN connection. Throughout, we assume a cosmology with 
$H_0=65$ kms$^{-1}$Mpc$^{-1}$, $\Omega_m=0.3$ and $\Omega_\Lambda=0.7$

\section{The Optical Light Curve}
\label{sec:lightcurve}

The optical light curves of GRB~990712 in different bands all seemed to decay 
as $t^{-1}$, with a single power-law decay plus a constant host contribution 
providing the best fit to the data (S00). In these fits, however, the host
magnitude was a free parameter in each band. Adding a supernova component of
type SN1998bw at the appropriate redshift did not change the derived decay 
rate. As noted by S00, a decay rate of $\alpha \gtrsim -1$, continuing indefinitely, 
would imply an infinite energy release in the burst. Therefore, light curves
decaying with $\alpha=-1$ or slower are required to break or bend.

Here, we use the ground based measurements of the host magnitudes, $V=22.40\pm 0.08$ 
and $R=21.91\pm 0.04$, as determined by Hjorth et~al.\ (2000), reducing the number 
of free parameters. We concentrate on the $V$-band as the host is relatively faint there. 
In Fig.\ 2 of S00, the three data points on July 14 may indicate a break in the $V$-band.
We therefore adopted all $V$-band data from S00 and re-reduced the subset of all publically 
available ESO $V$-band data for consistency and independence.
We also included the {\em HST} data taken 47.7 days 
after the burst (Fruchter et~al.\ 2000), assuming a power law spectrum to convert the
STIS magnitude to a $V$-band magnitude. Our $V$-band data, measured using aperture 
photometry, is presented in Table 1. The $R$-band data is taken unmodified from S00 
as the re-reduction of the $V$-band data turned out to be unnecessary.
The $I$-band data points are too few and far between to allow a reliable fit, in
addition to an unknown host magnitude in that band. 

Our $V$-band light curve is shown in Fig.\ 1, with the host magnitude subtracted. Also 
plotted is a broken power law fit to the light curve prior to day 7. We find that the 
initial light curve decay has a power law index of $\alpha_1=-0.83\pm 0.03$, while 
$\alpha_2=-3.06\pm 1.28$, with the break occurring at $t_b=1.61\pm 0.19$ days 
($\chi^2_4=0.434$, where $\chi^2_{\rm DOF}=\chi^2/{\rm DOF}$, is the reduced $\chi^2$ 
of the fit).
We note that $\alpha_1$ is significantly larger than the slope reported by S00. 
An unbroken power law fit to the entire data set results in $\alpha=-0.82\pm 0.03$,
but the fit is considerably worse ($\chi^2_9=1.73$). To estimate 
the effect of the uncertainty in the host magnitude on the decay rate we subtracted several 
host magnitudes from the light curve. Varying the magnitude in steps from $V=22.32$ to $22.55$, 
resulted in an increase in $\alpha_1$ from $-0.85\pm 0.04$ to $-0.79\pm0.04$. 

The post-break decay slope, $\alpha_2$, is very sensitive to the break time. If we fix the 
break time at $t_b=1.5$ days, then $\alpha_2=2.42\pm0.52$ ($\chi^2_5=0.734$). The evidence 
for the break is not very strong, however, as it hinges mostly on one data point (July 14.787), 
the reliability of which we are unable to verify (adopted from S00). Leaving that point out,
a single power law fit to the entire data set gives $\alpha=-0.81\pm 0.03$, ($\chi^2_8=0.696$). 

A break at $t_b\approx1.5$ days would be in an agreement with the BL00 interpretation of 
the polarization data, that the observed early light curve results from a sideways expanding 
jet. The light curve in that model should steepen by $\Delta \alpha=1-\alpha_1/3=1.28\pm 0.01$, 
quite consistent with the observed steepening of $2.23\pm 1.28$.

We show the $R$-band host subtracted light curve in Fig.\ 1. It differs from the $V$-band 
light curve in two important ways. Firstly, 
it is decaying somewhat faster than the $V$-band. A fit to the data from the first 4 days 
results in $\alpha= -0.94\pm 0.02$ ($\chi^2_{11}=1.39$), that is consistent with the S00 result. 
Again, using host magnitudes in the interval $21.74-21.95$, we find increasing decay rates 
in the range $-0.98\pm 0.02$ to $-0.91\pm 0.02$, respectively. Restricting the fit to the 
first 1.5 days does not affect this result. The difference between the early $V$-band decay rate 
and the $R$-band decay rate is significant and is not expected in fireball models, but we have no 
plausible explanation for it. 
Secondly, the arguments given in the previous paragraphs, for why there should be a 
break in the $V$-band also apply here. No such break is, however, seen in the $R$-band data.

If the break at 1.5 days in the $V$-band is real, the late time light curve 
is seen to be dominated by a component that is first clearly detected at a burst age of 
about 7 days. Although no break is observed in the $R$-band, this late time component 
can also be clearly seen there as a single power law provides a bad fit to the entire $R$-band 
data set ($\alpha= -0.84\pm 0.01$ with $\chi^2_{16}=8.98$). This late time component 
appears brighter at earlier times in $R$ than in the $V$-band and this may be the reason 
for why a break is not detectable in $R$. 

It is simplest to interpret this late time component as being due to a supernova that 
rises to a maximum in $V$ of about 23-24 mag at a burst age of 1 to 3 weeks. We also 
show in Fig.\ 1, the late time light curve of SN1998bw at the redshift of GRB~990712, 
$z=0.434$ (Hjorth et~al.\ 2000; Vreeswijk et~al.\ 2000). The GRB~990712 light curve 
has a reasonably good resemblance to the light curve of SN1998bw, being almost 
equally bright in the $R$-band and somewhat brighter in the $V$-band.
We emphasize that due to insufficient data coverage in the $V$-band, we have not
attempted to fit a broken power law light curve plus a supernova component to the data. 
We overplot the SN1998bw light curve simply to illustrate the similarity of the GRB~990712 
late light curve to it. In addition, it is unknown if SN998bw is typical of GRB associated 
supernovae. If GRB associated supernovae turn out to be standard candles, fitting late 
time afterglow light curves to such a standard would provide an independent measurement 
of the cosmological constant.

\section{Discussion}
\label{sec:disc}
In re-analyzing the $V$-band light curve of GRB~990712, we have benefited from 
the measured host magnitudes, reducing the number of free parameters in our fits. 
The lack of $V$-band data from day 1 to day 7, makes it difficult to quantify the 
significance of a break in the light curve. The strongest evidence for a break may 
come from the light curve and the polarization data {\em together}, as the same physical 
model then accounts for both measurements, implying that that the early 
light curve is produced in a collimated outflow. 

The $R$-band light curve, although not 
showing a break and therefore not a clear signature of a jet, does reveal a late time 
light curve behavior similar to the SN1998bw light curve. A supernova-like component then 
becomes a necessary ingredient to explain the late time light curves. The argument can also be
reversed, because if we accept the late time behavior in the $R$-band as being due 
to a supernova, we should expect similar behavior in the $V$-band. A break in the $V$-band 
light curve is then demanded by the data as otherwise an unbroken power law fits the light 
curve. In addition, energetics requires breaks in the light curves because the decay 
rates in both $V$ and $R$ are significantly greater than $-1$. 

In S00's fit including a supernova component, it was assumed that it had properties identical 
to SN1998bw. This is a very strong assumption and, as noted by S00, is hard to justify by 
current statistics of GRB/SN associations. We choose not to constrain our fit by SN1998bw,
but rather use it to demonstrate the possibility that a supernova may have been present in GRB~990712. 
If the difference between the light curves of GRB~990712 and SN1998bw is real, 
it may be due to differences in kinetic energy, composition, explosion geometry or in the 
properties of the local environment (e.g.\ Nakamura et~al.\ 2000; Nomoto et~al.\ 2000; 
Sollerman et~al.\ 2000). 

The case of GRB~990712 provides independent evidence from two different sets of measurements 
for a collimated outflow in a GRB, i.e.\ the polarization measurements (interpreted by BL00) 
and the light curve properties (this Letter). The data also shows a strong signature of a 
supernova like component, especially in the $R$-band, that in turn requires a break in 
the early light curve.
It is crucial that as complete time coverage as possible be attempted for future optical 
afterglows for at least a full month, to be able to discern the properties advocated in this 
Letter. The first few days are most demanding as a break in the light curve is 
expected to fall within this burst age. It is also important to follow the late time
behavior closely to look for a possible supernova accompanying the burst.  

\acknowledgements
This work was supported by the Icelandic Research Council,
the University of Iceland Research Fund, and the Danish Natural Science
Research Council (SNF). SH would like to acknowledge support from NASA 
grant NAG5-9364. We thank the anonymous referee for a number of suggestions
that improved the presentation.



\begin{figure}[t]
\epsscale{0.3}
\epsscale{1.}
\plottwo{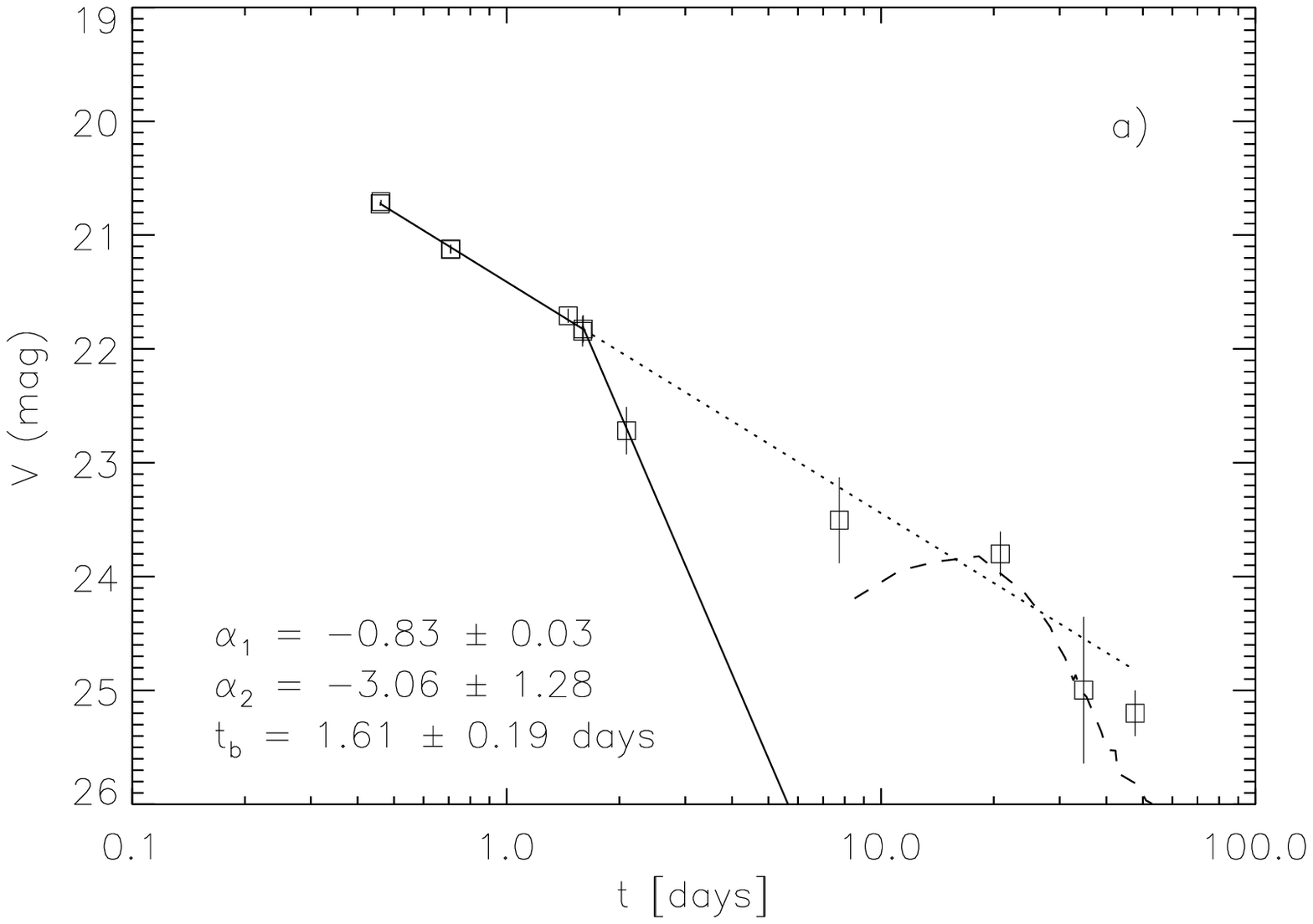}{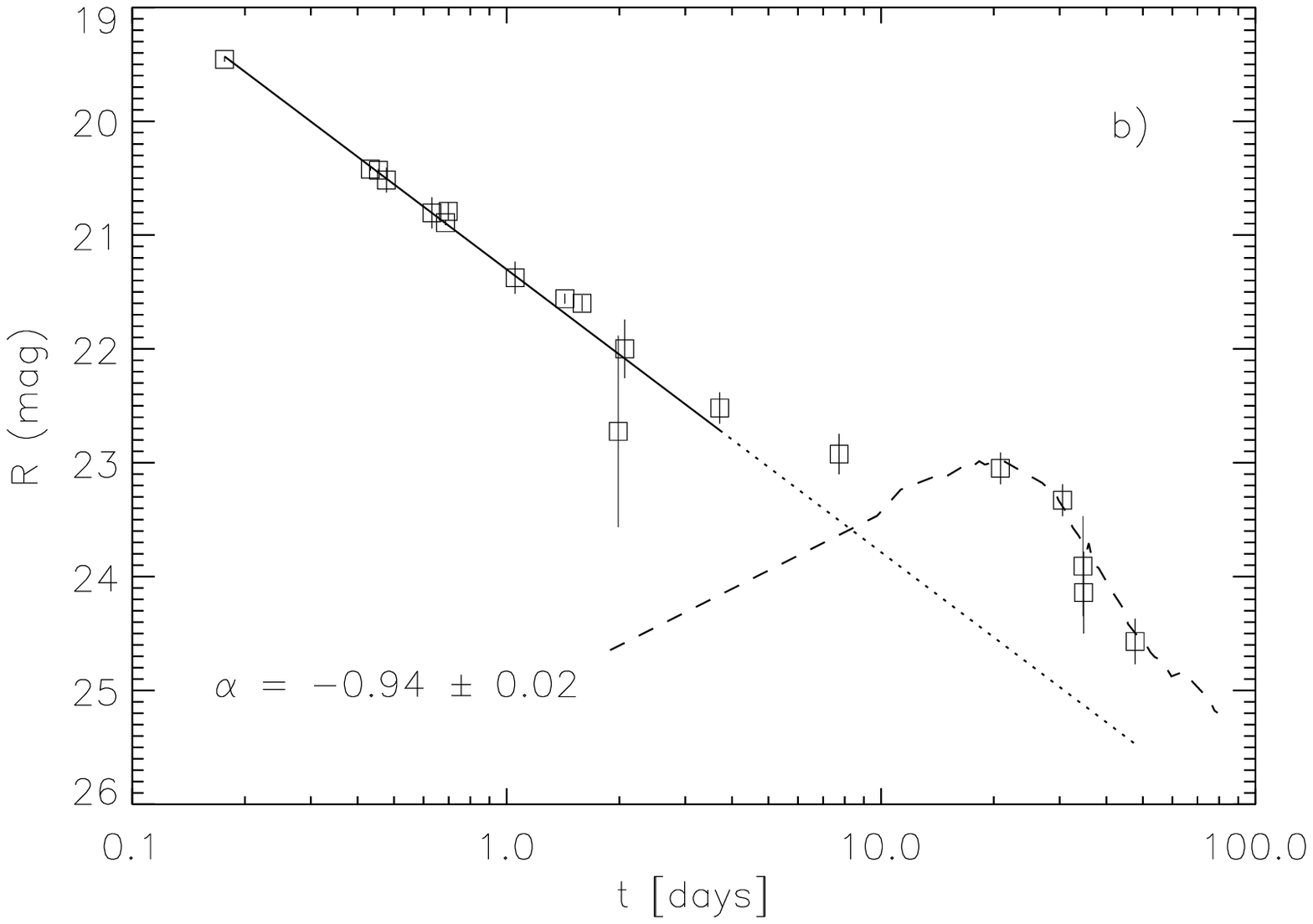}
\caption{
The left panel shows the $V$-band light curve of GRB~990712. Note the break after about 
1.6 days and the prominent supernova component. The light curve of SN1998bw at $z=0.434$ 
is shown dashed. A rather wide gap is in the data at a most crucial interval between 1 and 
7 days. The dotted line is an extrapolation of the early light curve. The right panel 
shows the $R$-band light curve, again with the SN1998bw light curve superimposed. 
A solid line shows a fit to the first 4 days. There is no evidence for a break in 
this case, but the supernova component rises well above the extrapolated power law fit. 
}
\label{fig:fig1}
\end{figure}


 
\begin{deluxetable}{lcr}
\footnotesize
\tablecaption{$V$-Band Photometry of GRB~990712\label{tab:obslog}}
\tablewidth{0pt}
\tablehead{
\colhead{Day (1999 UT)} & \colhead{Telescope} & \colhead{$V$ magnitude}
}
\startdata

Jul 13.156   & VLT  &    20.515 $\pm$ 0.013\\
Jul 13.158   & VLT  &    20.500 $\pm$ 0.012\\
Jul 13.405   & VLT  &    20.834 $\pm$ 0.026\\
Jul 13.406   & VLT  &    20.831 $\pm$ 0.029\\
Jul 14.157   & VLT  &    21.248 $\pm$ 0.040\\
Jul 14.292   & NTT  &    21.335 $\pm$ 0.083\\
Jul 14.298   & NTT  &    21.322 $\pm$ 0.076\\
Jul 14.787   & SAAO &    21.795 $\pm$ 0.089\tablenotemark{a}\\
Jul 20.433   & NTT  &    22.065 $\pm$ 0.100\tablenotemark{a}\\
Aug 02.511   & AAT  &    22.136 $\pm$ 0.042\tablenotemark{a}\\
Aug 16.471   & AAT  &    22.305 $\pm$ 0.054\tablenotemark{a}\\
Aug 29.403   & HST  &    25.25  $\pm$ 0.2\tablenotemark{b}

\enddata
\tablenotetext{a}{Taken from  Sahu et~al.\ (2000).}
\tablenotetext{b}{OT magnitude}

\end{deluxetable}


\begin{thebibliography}{}
\bibitem[2000]{}
Bj\"ornsson, G.\ \& Lindfors, E.\ J.\ 2000, ApJ, 541, L55 (BL00)

\bibitem[2000]{}
Bloom, J.\ S.\ et~al.\ 1999, Nature, 401, 453

\bibitem[2000]{}
Castro-Tirado, A.\ J.\ et~al.\ 1999 Science, 283, 2069

\bibitem[2000]{}
Dar, A.\ \& De R\'ujula, A., 2000, (astro-ph/0008474) 

\bibitem[2000]{}
Fruchter, A., et~al.\ 2000, GCN 752

\bibitem[2000]{}
Galama, T.\ J., et~al.\ 1998, Nature, 395, 670

\bibitem[2000]{}
Galama, T.\ J., et~al.\ 2000, \apj, 536, 185

\bibitem[2000]{}
Germany, L.\ M., Reiss, D.\ J., Sadler, E.\ M., Schmidt, B.\ P.\ \& Stubbs, C.\ W.\
2000, \apj, 533, 320

\bibitem[2000]{}
Harrison F.\ A., et~al.\ 1999, ApJ, 523, L121

\bibitem[2000]{}
Hjorth, J., Holland, S., Courbin, F., Dar, A., Olsen, L.\ F.\ \& Scodeggio, M.\ 2000, 
\apj, 534, L147; 539, L75 

\bibitem[2000]{}
Holland, S., Bj\"ornsson, G., Hjorth, J.\ \& Thomsen, B.\ 2000, A\&A, 364, 467

\bibitem[2000]{}
Holland, S., et~al.\ 2001, A\&A, in press, (astro-ph/0103044)

\bibitem[2000]{}
Israel, G.\ L., et~al.\ 1999, A\&A, 348, L5

\bibitem[2000]{}
Jaunsen, A.\ O., et~al.\ 2001, ApJ, 546, 127

\bibitem[2000]{}
Jensen, B.\ L., et~al.\ 2000, A\&A, in press (astro-ph/0005609)

\bibitem[2000]{}
Kulkarni, S.\ R., et~al.\ 1999, Nature, 398, 389

\bibitem[2000]{}
MacFadyen, A.\ I.\ \& Woosley, S.\ E.\ 1999, ApJ, 524, 262

\bibitem[2000]{}
Nakamura, T., Mazzali, P.\ A., Nomoto, K.\ \& Iwamoto, K.\ 2000, \apj, in press
(astro-ph/0007010)

\bibitem[2000]{}
Nomoto, K.\ et~al., 2000, to appear in Supernovae and Gamma Ray Bursts, ed.\ M.\ Livio et~al.\
(astro-ph/0003077)

\bibitem[2000]{}
Paczy\'nski, B.\ 1998, 494, L45

\bibitem[2000]{}
Reichart, D.\ E., 1999, \apjl, 521, L111

\bibitem[2000]{}
Rol, E., et~al.\ ApJ, 544, 707

\bibitem[2000]{}
Sahu, K.\ C., et~al.\ 2000, \apj, 540, 74 (S00)

\bibitem[2000]{}
Sari, R., Piran, T.\ \& Narayan, R.\ 1998, ApJ, 497, L17

\bibitem[2000]{}
Sari, R., Piran, T.\ \& Halpern, J.\ P.\ 1999, ApJ, 519, L17

\bibitem[2000]{}
Sollerman, J., Kozma, C., Fransson, C., Leibundgut, B., Lundqvist, P., Ryde, F.\
\& Woudt, P.\ 2000, ApJ, 537, L127

\bibitem[2000]{}
Stanek K.\ Z., Garnavich, P.\ M., Kaluzny, J., Pych, W.\ \& Thompson, I.\ 1999, ApJ, 522, L39

\bibitem[2000]{}
Turatto, M., et~al.\ 2000, ApJ, 534, L57

\bibitem[2000]{}
Vreeswijk, P.\ M., et~al.\ 2000, ApJ, 546, 672

\bibitem[2000]{}
Woosley, S.\ E.\ 1993, ApJ, 405, 273

\end{thebibliography}
\end{document}